\documentclass[conference]{IEEEtran}
\IEEEoverridecommandlockouts
\usepackage{cite}
\usepackage{amsmath,amssymb,amsfonts}
\usepackage{algorithmic}
\usepackage{algorithm}
\usepackage{bm}
\usepackage{graphicx}
\usepackage{textcomp}
\usepackage{xcolor}
\usepackage{array}
\usepackage{textcomp}
\usepackage{stfloats}
\usepackage{subcaption}
\usepackage{url}
\usepackage{verbatim}
\usepackage{cite}
\usepackage{booktabs}
\def\BibTeX{{\rm B\kern-.05em{\sc i\kern-.025em b}\kern-.08em
    T\kern-.1667em\lower.7ex\hbox{E}\kern-.125emX}}
\begin{document}

\title{Doppler-Resilient Rydberg Atomic Receiver for High-Dynamic Communication Networks via Adaptive Local Oscillator Tracking\\

}
\author{Yiyue Xiang\textsuperscript{1}, 
	Jianxiong Pan\textsuperscript{1},
	Qiaolin Ouyang\textsuperscript{2},
	Bichen Kang\textsuperscript{1},
	Bin Qi\textsuperscript{2}, 
	Neng Ye\textsuperscript{1} \\
\IEEEauthorblockA{\textsuperscript{1} School of Cyberspace Science and Technology, Beijing Institute of Technology, Beijing, China}
\IEEEauthorblockA{\textsuperscript{2} School of Information and Electronics, Beijing Institute of Technology, Beijing, China}
}
	

\maketitle

\begin{abstract}
Rydberg atomic receiver has emerged as promising candidate for next-generation wireless communication, due to the exceptional sensitivity and ability to overcome the physical limitations of traditional radio frequency antennas. Utilizing the resonant response of atomic energy levels for signal detection, Rydberg atomic receiver is inherently confined to a narrow instantaneous bandwidth. However, in high-mobility scenarios such as satellite communications, the severe Doppler effect induces carrier frequency offsets, which drive the signal beyond the instantaneous bandwidth and result in severe distortion. In this paper, we propose an adaptive local oscillator (LO) tracking Rydberg atomic receiver architecture designed to lock high-dynamic signals within the effective atomic response bandwidth. By employing a cross-product automatic frequency control (CPAFC) algorithm, the system dynamically estimates the instantaneous frequency offset, generates a corresponding error control signal, and adjusts the LO frequency through a feedback loop. Consequently, the intermediate frequency signal can always be locked close to the center of the atomic response bandwidth regardless of dynamics. Simulation results show that the proposed architecture significantly outperforms existing Rydberg atomic receiver, effectively alleviating performance degradation in high-dynamic environments.
\end{abstract}


\section{Introduction}
Wireless communication is undergoing a persistent evolution towards broader frequency bands and multi-platform integration. With this trend, the Rydberg atomic receiver has emerged as a revolutionary quantum technology. Leveraging the strong dipole coupling of Rydberg atoms to incident electric fields, this technology enables the direct detection of signals ranging from direct current to terahertz (THz) frequencies \cite{schlossberger2024rydberg}. Furthermore, it offers inherent advantages including high sensitivity, wavelength-independence, intrinsic SI-traceability, broad spectral tunability, and considerable potential for wideband coverages \cite{schlossberger2024rydberg,meyer2020assessment}. Owing to these attributes, Rydberg atomic receiver holds compelling promise for the next-generation applications, e.g., THz communications, wideband spectrum sensing, and satellite networks.  

Rydberg atomic receiver is founded on quantum coherent effects, primarily electromagnetically induced transparency (EIT) and the Autler-Townes (AT) splitting effects \cite{anderson2020rydberg}. In this approach, an optical readout scheme maps the frequency and amplitude modulation of an incident radio-frequency (RF) electric field onto the transmitted intensity of a probe laser passing through a vapor cell filled with alkali atoms. This enables signal reception without the need for traditional antennas, thereby eliminating their wavelength-dependent constraints. Although its exceptional sensitivity and frequency agility have been established both theoretically and experimentally \cite{jing2020atomic,cai2023high,wen2024rydberg}, the underlying physical mechanism imposes a critical constraint that the instantaneous bandwidth of RF signal to be detected is inherently limited by the finite linewidth of the involved Rydberg transition. This typically confines the usable bandwidth to the megahertz (MHz) scale or narrower \cite{holloway2019detecting,fancher2021rydberg}, constituting a critical bottleneck for processing wideband or rapidly frequency-varying signals.

To accommodate the demodulation requirements of dominant digital modulation schemes, e.g., QPSK, the majority of existing Rydberg atomic receivers employ a heterodyne architecture relying on a fixed local oscillator (LO) \cite{jing2020atomic,wu2023linear,gong2025rydberg}. This approach down-converts the RF signal to an intermediate frequency (IF) that falls within the atomic instantaneous bandwidth. While this architecture performs well in static or quasi-static channels, it faces significant challenges in high-mobility scenarios such as satellites communication, where the severe Doppler effect induces rapid and substantial carrier frequency offsets. With a fixed LO, these offsets cause the down-converted IF signal to drift beyond the flat region of the effective atomic response, which leads to severe signal distortion and results in significant performance deterioration or even communication outage. In the current literature, effective solutions for dynamic frequency tracking and compensation in such bandwidth-constrained quantum receivers still remain largely absent.

In this paper, we develop a Doppler-resilient Rydberg atomic receiver for high-dynamic communication networks, where an adaptive LO tracking architecture is proposed for the first time. We identify that, unlike traditional RF receiver, Rydberg atomic receiver is subject to distinctive impacts of Doppler shifts. On this basis, a feedback control loop tailed to Rydberg atomic receiver, driven by a cross-product automatic frequency control (CPAFC) algorithm, is developed to estimate the instantaneous Doppler shifts and adjust the LO frequency dynamically. 

The remainder of this paper is organized as follows. Section II introduces the system model of LO-based Rydberg atomic receiver. In Section III, we analyze the impacts of Doppler shifts on Rydberg atomic receiver and develop an adaptive LO tracking receiver architecture. Section IV presents the simulation results and Section V concludes this paper.

\section{System Model of LO-Based Rydberg Atomic Receiver}
\label{sec:system_model}

\subsection{System Overview and Operating Principle}
The structure of LO-based Rydberg atomic receiver system, depicted in Fig.~1(a), functionally parallels a classical heterodyne receiver but replaces the RF front-end components, e.g., antenna, mixer and filter,  with a quantum sensor. It primarily consists of a vapor cell filled with alkali atoms, a strong LO, a photodetector, and two counter-propagating lasers.

\begin{figure}[!t]
	\centering
	\includegraphics[width=\linewidth]{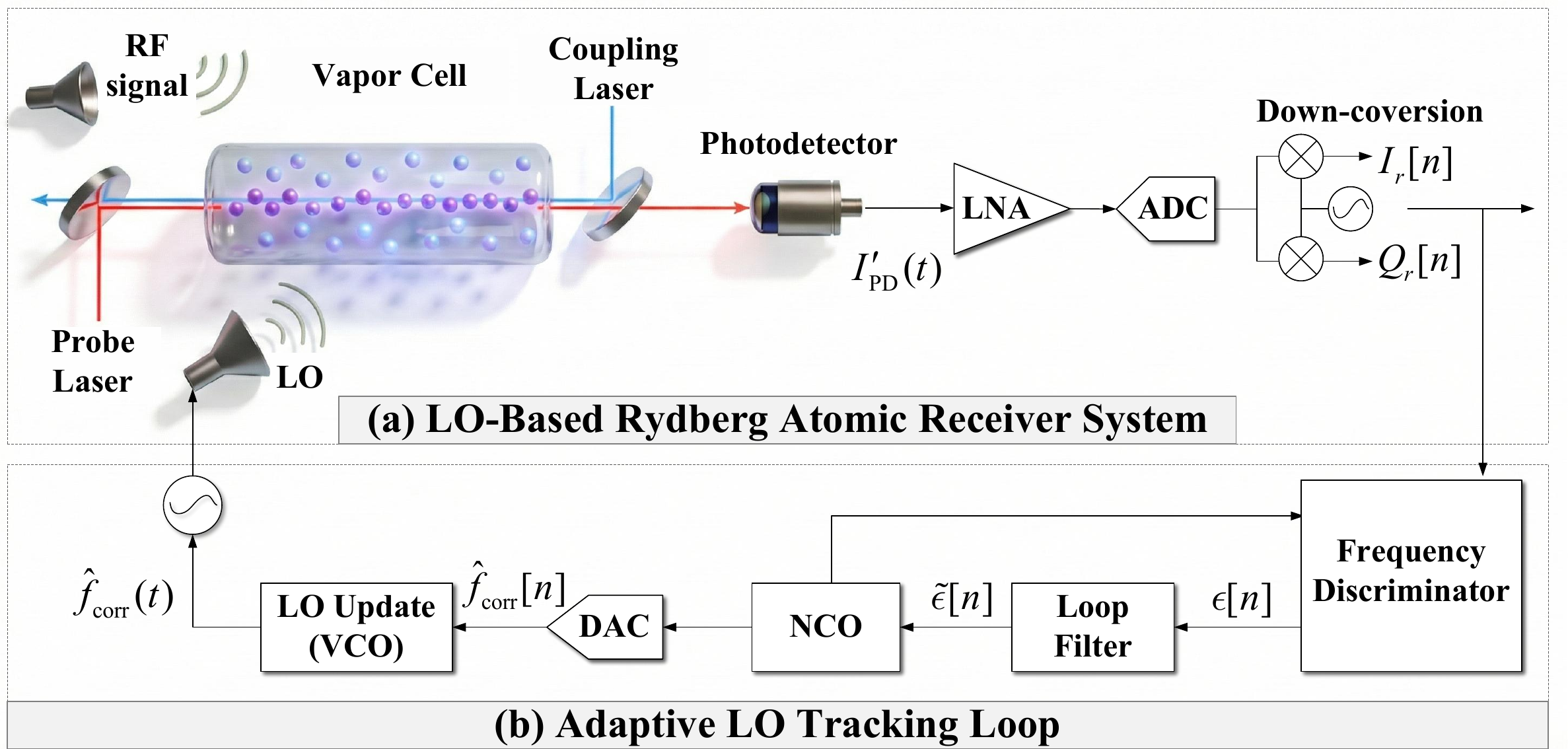}
	\caption{Proposed adaptive LO tracking Rydberg atomic receiver architecture.}
	\label{system_model}
\end{figure}

The operating principle of LO-Based Rydberg atomic receiver is to transduce RF field into an optical signal via the coherent response of the prepared Rydberg atomic system. 
Typically, consider a four-level ladder-type atomic system, which involves a ground state $|g\rangle$, an excited state $|e\rangle$, and two Rydberg states $|r_1\rangle$ and $|r_2\rangle$. 
A probe laser and a coupling laser drive the transitions $|g\rangle \rightarrow |e\rangle$ and $|e\rangle \rightarrow |r_1\rangle$, respectively, establishing the electromagnetically induced transparency (EIT) condition for the probe.
An incident RF field and a strong LO field jointly couple the transition $|r_1\rangle \rightarrow |r_2\rangle$. The LO induces a dominant Autler-Townes (AT) splitting of the EIT resonance, effectively biasing the atomic system. The weaker RF signal then perturbs this bias and alters the transmission of probe laser, amplitude-modulating the atomic response at the beat frequency. Then, this amplitude-modulated optical transmission is converted into a photocurrent, yielding an electrical IF signal that carries the original RF modulation.


%
%

\subsection{Signal Model From Transmitter to Receiver}
At the transmitter, the complex baseband signal can be expressed as $ x_b(t) = \sum_k a_k g(t - kT_s) $, where \( a_k \) denotes complex symbols from a modulation alphabet, $ g(t) $ is the pulse‑shaping filter, and $ T_s $ is the symbol period. Equivalently, the baseband signal can be written in the form of \( x_b(t) = A_b(t) e^{j\phi_s(t)} \), where $A_b(t) $ and $ \phi_s(t) $ are the time‑varying amplitude and phase that carry the modulation, respectively. By up‑converting the baseband signal to the carrier frequency, we can obtain the transmitted RF signal as
\begin{equation}
	x(t) = \Re\left\{ x_b(t) e^{j2\pi f_c t} \right\}
	= A_{b}\sqrt{P} \cos\bigl(2\pi f_c t + \phi_s\bigr),
\end{equation}
where $ f_c $ is the carrier frequency of the RF signal.
For notation simplicity, we omit $t$ in $A_b(t)$ and $\phi_s(t)$ in the following.

After propagating through the wireless channel, the RF signal at the receiver can be represented as $ y(t) = \sum_l h_l(t) x(t-\tau_l) $, where $ h_l(t) $ and $\tau_l$ are the complex gain and delay of the $ l $-th path, respectively. Under the narrowband assumption (which will be discussed later), the received signal simplifies to $ y(t) = h(t) x(t) $. Rewriting the received signal as an electric field incident on the atomic vapor cell, we have
\begin{equation}
	E_{\text{RF}}(t) = A_{\text{RF}} \cos\bigl(2\pi f_c t + \phi_s\bigr),
	\label{eq:Erf_simple}
\end{equation}
where $ A_{\text{RF}} $ now incorporates both the original transmission amplitude and the channel gain $ |h(t)| $.

The LO provides a strong, monochromatic reference field
\begin{equation}
	E_{\rm LO}(t) = A_{\rm LO} \cos(2\pi f_{\rm LO} t + \phi_{\rm LO}),
	\label{eq:elo}
\end{equation}
where $f_{\rm LO}$, $A_{\rm LO}$, and $\phi_{\rm LO}$ denote the frequency, amplitude, and phase of the LO field, respectively.

The total field $E_{\rm tot}(t) = E_{\rm RF}(t) + E_{\rm LO}(t)$ drives the transition $|r_1\rangle \rightarrow |r_2\rangle$. In the typical operating regime where $A_{\rm LO} \gg A_{\rm RF}$, the atomic system behaves as a quantum mixer and the total field is derived as
\begin{equation}
	\begin{split}
		&E_{\rm tot}(t)\approx  \\
		&\left[A_{\rm LO} + A_{\rm RF}\cos(2\pi f_{\rm IF}t + \Delta\phi)\right]\cos(2\pi f_{\rm LO} t + \phi_{\rm LO})
	\end{split}
\end{equation}
where $f_{\rm IF} = |f_c - f_{\rm LO}|$ is the designed intermediate frequency, and $ \Delta\phi=|\phi_s-\phi_{\rm LO}|$. 

For the atomic system, it is the low-frequency beat component, i.e., $A_{\rm LO} + A_{\rm RF}\cos(2\pi f_{\rm IF}t + \Delta\phi)$, that modulates the atomic population and consequently the probe transmission. 



\subsection{Quantum Mechanisms and Optical Readout}
The response of the atomic system is governed by the Lindblad master equation \cite{auzinsh2010optically}, i.e.,
\begin{equation}
	\dot{\bm \rho} = -\frac{i}{\hbar}[\hat{\bm H}, \bm{\rho}] + \mathcal{L}(\bm \rho),
\end{equation}
where $\mathcal{L}(\rho)$ accounts for relaxation processes such as spontaneous emission and dephasing with rate $\Gamma$. The total Hamiltonian $\hat{\bm H} = \hat{\bm H}_{\rm atom} + \hat{\bm H}_{\rm laser} + \hat{\bm H}_{\rm field}$ governs the atomic system dynamics, where $\hat{\bm H}_{\rm atom}$ and $\hat{\bm H}_{\rm laser}$ describe the unperturbed atomic energy levels and the laser-atom interactions, respectively. The field interaction Hamiltonian $\hat{\bm H}_{\rm field}$ is derived from the Rabi frequency of the totoal field $\Omega_{\rm tot}(t) \propto A_{\rm LO} + A_{\rm RF}\cos(2\pi f_{\rm IF}t + \Delta\phi)$ \footnote{Due to space limitation, this paper omits the detailed derivation of some simple or unnecessary formulas. If needed, please refer to \cite{chen2025harnessing,gong2025rydberg}.}.


The solution for $\rho_{21}(t)$ in the density matrix $\bm{\rho}$ yields the probe laser susceptibility. Then, the intensity of the probe laser measured by the photodetector can be calculated by the imaginary part of the susceptibility, i.e., $P_{\rm out}(t) \propto \text{Im}\left\{\rho_{21}(t)\right\}$.

Based on the typical solution and approximation of $\rho_{21}(t)$ \cite{chen2025harnessing}, $P_{\rm out}(t)$ can be further represented as 
\begin{equation}
	P_{\rm out}(t) = {\bar P}_{0} + \kappa\cos (2\pi {f_{{\rm{IF}}}}t + \Delta\phi),
\end{equation}
where ${\bar P}_{0}$ is the average intensity of the probe laser and $\kappa$ is the key intrinsic coefficient of atomic response.

By subtracting the steady state response output ${\bar P}_{0}$, we can finally obtain the measurement of RF signal as
\begin{equation}
	I_{\rm PD}(t) \propto \alpha \kappa \cos(2\pi f_{\rm IF}t + \Delta\phi),
\end{equation}
where $\alpha$ denotes the photodetector responsivity. $I_{\rm PD}(t)$ is then amplified, digitized, and down-converted to obtain the baseband signal $r(t)$.

Note that system models in existing studies typically assume that $f_{\rm LO}$ is detuned slightly from $f_c$ by a fixed offset, ensuring the beat component modulating the atomic response is maintained at an optimal IF. This is a critical requirement for Rydberg atomic receiver.  Once the IF signal drifts away from the flat region of the atomic response, typically within $10$ MHz \cite{holloway2019detecting,fancher2021rydberg}, performance deteriorates drastically. This physical constraint is the primary limitation for processing high-dynamic signals, as discussed in the following section.

\section{Proposed Adaptive LO Tracking for Doppler-Resilient Receiver}
In this section, we identify the impact of Doppler shifts on Rydberg atomic receiver in high-mobility scenarios. On this basis, we present our proposed adaptive LO tracking architecture and the corresponding LO tracking algorithm.

\subsection{Impacts of Doppler Shifts on Rydberg Atomic Receiver}
\label{subsec:doppler_impact}

In high-dynamic communication networks, such as low earth orbit (LEO) satellite-terrestrial links, the relative motion between the transmitter and receiver may induce a significant, time-varying Doppler shift. This section analyzes the detrimental effects of such Doppler shifts on the fixed LO Rydberg atomic receiver architecture.

Let the time-varying Doppler shift be denoted as $f_d(t)$. The instantaneous frequency of the received RF signal becomes
\begin{equation}
	f_c'(t) = f_c + f_d(t).
	\label{eq:doppler_carrier}
\end{equation}

Under the conventional architecture, $f_{\rm LO}$ remains constant. Thus, the instantaneous intermediate frequency drifts away from its design value, becoming a time-dependent variable
\begin{equation}
	f'_{\rm IF}(t) = |f_c'(t) - f_{\rm LO}| \triangleq f_{\rm IF} + f_d(t),
	\label{eq:if_shifted}
\end{equation}
where $\triangleq$ holds for a reasonable assumption that $f_c > f_{\rm LO}$ and a defined sign convention for $f_d(t)$. 

Following the process described in Section~\ref{sec:system_model}, the photocurrent signal becomes
\begin{equation}
	I'_{\rm PD}(t) \propto \alpha \kappa \, \cos(2\pi f'_{\rm IF}(t) t + \Delta\phi),
	\label{eq:I_pd_doppler}
\end{equation}
and we denote the corresponding baseband signal as $r^{\prime}(t)$.

The drifting IF $f'_{\rm IF}(t)$, leads to the following two primary degradation mechanisms for the Rydberg atomic receiver.

First, the accumulated phase error due to the frequency offset, i.e., $\theta_e(t) = 2\pi \int_0^t f_d(\tau) d\tau$, causes a continuous rotation of the demodulated constellation. 
For phase-modulated signals like QPSK, this requires phase tracking or correction at the demodulator, a challenge also faced by conventional electronic receivers.

Second, the drifting IF may fall outside the narrow atomic instantaneous response bandwidth, which constitutes the distinct and critical challenge for the Rydberg atomic receiver. This can be described by an atomic equivalent channel response, defined as $H_{\rm a}(f)$, which acts on the IF signal. Typically, $H_{\rm a}(f)$ is a bandpass function centered at the nominal IF $f_{\rm IF}$, with a very limited instantaneous bandwidth defined as $B_{\rm a}$, often on the order of few MHz or even hundreds of kHz. Primarily constrained by the intrinsic linewidth of the atomic transition, its magnitude response can be approximated by a Lorentzian lineshape as
\begin{equation}
	|H_a(f)| \approx \frac{1}{\sqrt{1 + \left(\frac{2|f - f_{\rm IF}|}{B_{\rm a}}\right)^2}}.
	\label{eq:atom_response_doppler}
\end{equation}
The consequence of the drifting $f'_{\rm IF}(t)$ is that the signal spectrum is no longer aligned with the center of $H_a(f)$. When the instantaneous frequency deviation exceeds half the atomic bandwidth, i.e.,
\begin{equation}
	|f'_{\rm IF}(t) - f_{\rm IF}| = |f_d(t)| > B_{\rm a}/2,
	\label{eq:bandwidth_condition}
\end{equation}
the signal power experiences severe attenuation due to the roll-off of $|H_a(f)|$.


The analysis above demonstrates that a fixed LO architecture is fundamentally inadequate for high-dynamic scenarios. The problem is exacerbated at higher carrier frequencies (e.g., THz bands) or with higher platform velocities. 
For instance, in LEO satellite scenarios, the relative velocity and acceleration can reach $\sim 7 km/s$ and $\sim 81.6 m/s^2$, respectively \cite{tr38811}. Under such dynamics, carrier frequencies on the order of hundred-GHz will induce Doppler shifts of several MHz per second.

Therefore, maintaining the effective intermediate frequency $f'_{\rm IF}(t)$ within the optimal operating range, specifically within $\pm B_{\rm a}/2$ of the nominal $f_{\rm IF}$ is critical. This requirement motivates the design of an adaptive LO tracking system, which aims to dynamically adjust $f_{\rm LO}$ to compensate for $f_d(t)$, thereby stabilizing the IF experienced by the atoms. 


\subsection{Proposed Adaptive LO Tracking Architecture}
\label{subsec:architecture}

To address the aforementioned challenges, we propose a adaptive LO tracking architecture, as depicted in Fig.~1. The core concept is to shift the burden of frequency compensation from the digital demodulation stage to the physical front-end. By employing a frequency-locked loop that estimates the instantaneous Doppler-induced offset and feeds back a correction signal, the hardware LO frequency is dynamically tuned to physically compensate for $f_d(t)$ before the RF signal is down-converted to the IF. 

The proposed architecture builds upon the LO-based system by integrating a feedback control loop shown in Fig.~1(b). The overall structure, consisting of both the classic chain and the proposed adaptive components, is elaborated as follows:


\paragraph{Analog-to-Digital Conversion and Down-mixing} 
The amplified IF signal is digitized by an analog-to-digital converter (ADC). The digital samples are then quadrature-down-mixed using a digital LO set to the nominal IF $f_{\rm IF}$. This yields a complex baseband signal, i.e., $r^{\prime}[n] = I_r[n] + jQ_r[n]$, where
\begin{equation}
	\begin{split}
		&I_r[n] = A\cos(2\pi\Delta f_e[n]nT+ \Delta\phi),\\
		&Q_r[n] = -A\sin(2\pi\Delta f_e[n]nT+ \Delta\phi)
	\end{split}
\end{equation} 
where $\Delta f_e[n]$ is the residual frequency offset, $T$ is the sample period, and $A$ is the amplitude.

\paragraph{Frequency Discriminator} The complex baseband signal in the form of $I_r[n]$ and $Q_r[n]$ are then fed into the core of the feedback loop, i.e., frequency discriminator. This module aims to extract an estimate of the instantaneous frequency error $\Delta f_e[n]$ at the $n$-th sample, defined as $\epsilon[n]$, from the phase progression between successive symbols, making $\epsilon[n] \approx \Delta f_e[n]$.

\paragraph{Loop Filtering} The raw error estimation $\epsilon[n]$ is then passed through a digital loop filter, typically a proportional-integral type. The filter suppresses high-frequency noise as well as outliers in the error signal, and defines the dynamic response involving bandwidth and damping of the overall control loop. Its output is a smoothed frequency control signal, denoted as $\tilde{\epsilon}[n]$, ensuring stable and convergent tracking.

\paragraph{Digital Phase Correction and Analog LO Frequency Update} The control signal $\tilde{\epsilon}[n]$ drives two parallel paths:
\begin{itemize}
	\item Digital phase correction: A numerically controlled oscillator (NCO) uses $\tilde{\epsilon}[n]$ to generate a complex signal that digitally de-rotates the baseband signal $r^{\prime}[n]$, providing immediate fine phase adjustment in the digital domain.
	\item Analog LO frequency update: $\tilde{\epsilon}[n]$ is integrated into a cumulative correction $\hat{f}_{\rm corr}[n]$, which is converted by a DAC into an analog tuning voltage for the voltage‑controlled oscillator (VCO). This physically shifts the LO frequency to $f^{\prime}_{\rm LO}(t) = f_{\rm LO} + \hat{f}_{\rm corr}(t)$, thereby realigning the down‑conversion process at the analog front‑end.
\end{itemize}

\paragraph{Closed-Loop Operation} The adjusted LO frequency $f^{\prime}_{\rm LO}(t)$ is fed back to the mixer in the analog front-end. This changes the effective down-conversion process, counter-acting the Doppler shift $f_d(t)$. The objective of the control loop is to nullify the effective IF drift, enforcing the condition
\begin{equation}
	f'_{\rm IF}(t) = (f_c + f_d(t)) - f^{\prime}_{\rm LO}(t) \approx f_{\rm IF}.
	\label{eq:tracking_objective}
\end{equation}
When the loop is locked, $\hat{f}_{\rm corr}(t) \approx f_d(t)$, and the signal for atomic system remains centered at the optimal IF, ensuring that the atoms always interact with a signal frequency that yields the optimal response, regardless of the external Doppler shift.


\subsection{Cross-Product-based LO Tracking Algorithm}
\label{subsec:algorithm}

The key component of the feedback loop is the frequency error estimation algorithm. We employ a cross-product automatic frequency control (CPAFC) algorithm, 
which processes the baseband signal $r^{\prime}[n]$ derived from the photodetector output after ADC and down-mixing. 


The CPAFC algorithm consists of the following steps:

\subsubsection{Modulation Wiping}
Since the received signal $r^{\prime}[n]$ contains unknown phase modulation $\Delta\phi$, direct frequency discrimination would be corrupted by phase jumps. To tackle this issue, we raise the signal to the $M$-th power that removes the modulation, and the resulting sample will be
\begin{equation}
	z[n] = \left( \frac{r^{\prime}[n]}{|r^{\prime}[n]|} \right)^M = e^{j(M\cdot2\pi \Delta f_e n T + M\Delta\phi)}.
	\label{eq:mod_wipe}
\end{equation}
Since $M\Delta\phi$ can be an integer multiple of $2\pi$, the modulation term vanishes, leaving only the carrier frequency offset term amplified by a factor of $M$, i.e.,
\begin{equation}
	z[n] \approx e^{j(M\cdot2\pi \Delta f_e n T)}.
	\label{eq:z_clean}
\end{equation}

\subsubsection{Frequency Discrimination}
The frequency error is estimated using a cross-product discriminator, which approximates the phase difference between consecutive samples $z[n]$ and $z[n-1]$. Define the cross-product as
\begin{equation}
	P_{\text{cross}}[n] = I_z[n]Q_z[n-1]-Q_z[n]I_z[n-1].
	\label{eq:cross_prod}
\end{equation}
where $I_z[n]$ and $Q_z[n]$ are the in-phase and quadrature of $z[n]$, respectively.
Substituting \eqref{eq:z_clean} and assuming small frequency changes between samples, we obtain
\begin{equation}
	P_{\text{cross}}[n] = \sin(M\cdot2\pi \Delta f_e T).
	\label{eq:cross_sin}
\end{equation}
For small arguments ($|M\cdot2\pi \Delta f_e T| \ll 1$), $\sin(x) \approx x$, leading to 
\begin{equation}
	P_{\text{cross}}[n] \approx M\cdot2\pi \Delta f_e T
\end{equation}
Thus, the raw frequency error estimate is obtained through the division by $M$ to normalize the error back to the original frequency scale, i.e.,
\begin{equation}
	\epsilon[n] = \frac{P_{\text{cross}}[n]}{M\cdot2\pi T_{\rm coh}} 
	\label{eq:raw_error}
\end{equation}
where $T_{\rm coh}$ is the coherent integration time, which is typically set to the symbol duration $T_{s}$. 

\subsubsection{Loop Filtering}
To smooth the error signal and provide appropriate loop dynamics, the error $\epsilon[n]$ is passed through a second-order digital loop filter. 
The filter implements a proportional-integral controller in the discrete domain, with a discrete transfer function of
\begin{equation}
	F(z) = \frac{p_1-p_2z^{-1}}{K(1-2z^{-1}+z^{-2})},
	\label{eq:pi_filter}
\end{equation}
where $p_1=2\zeta\omega_nT_s+\omega_n^2T_s^2$ and $p_2=-2\zeta\omega_nT_s$.
These coefficients are determined by the loop natural frequency $\omega_n$ and damping factor $\zeta$, designed to balance tracking speed and noise suppression. 
Thus, the filter output $\tilde{\epsilon}[n]$ is updated recursively as
\begin{equation}
	\tilde{\epsilon}[n] = 2\tilde{\epsilon}[n-1]-\tilde{\epsilon}[n-2] + \frac{p_1}{K} \epsilon[n] +\frac{p_2}{K} \epsilon[n-1].
	\label{eq:filter_update}
\end{equation}
where $K$ is the equivalent digital loop gain.

\subsubsection{LO Frequency Update}
The filter output $\tilde{\epsilon}[n]$ drives the frequency correction through dual complementary paths:
\begin{itemize}
	\item Digital phase correction: The control signal $\tilde{\epsilon}[n]$ directly sets the instantaneous frequency of a NCO. The resulting complex exponential $e^{-j 2\pi \tilde{\epsilon}[n] n T}$ is multiplied with the baseband signal $r^{\prime}[n]$, providing per‑sample de‑rotation in the digital domain.
	\item Analog LO frequency update: The control signal $\tilde{\epsilon}[n]$ is integrated to form a cumulative frequency correction $\hat{f}_{\text{corr}}[n] = \sum_{k=1}^{n} \tilde{\epsilon}[k]$. This term is converted to an analog voltage via a DAC and fed to the LO update module that tunes the VCO, shifting the physical LO frequency to $f^{\prime}_{\text{LO}}(t) = f_{\text{LO}} + \hat{f}_{\text{corr}}[n]$. This alters the subsequent intermediate frequency signal interacting with the atoms.
\end{itemize}


In steady state, the loop drives the error $\epsilon[n]$ toward zero, which implies $\Delta f_e \rightarrow 0$ and, consequently, $f^{\prime}_{\text{LO}}(t) \approx f_{\text{LO}} + f_d(t)$.
Thus, the combined action of both correction paths stabilizes the effective intermediate frequency at the nominal $f_{\text{IF}}$, keeping it within the atomic optimal response bandwidth.

The complete CPAFC-based adaptive LO tracking procedure is summarized in Algorithm~\ref{alg:cpafc}.

\begin{algorithm}[!t]
	\small
	\caption{CPAFC for Rydberg Atomic Receiver}
	\label{alg:cpafc}
	\begin{algorithmic}[1]
		\STATE \textbf{Input:} Baseband samples $r[n]$, symbol period $T_s$, modulation order $M$, loop filtering coefficients $p_1, p_2$.
		\STATE \textbf{Initialization:}  $z[0] \gets 0$, $\epsilon \gets 0$, $\tilde{\epsilon} \gets 0$, $\hat{f}_{\text{corr}} \gets 0$
		\FOR{each processing block}
		\STATE \% \textbf{Modulation Wiping:}
		\STATE \quad Normalize amplitude $\hat{r}[n] = r[n] / |r[n]$ 
		\STATE \quad Compute $M$-th power $z[n] = (\hat{r}[n])^M$ 
		\STATE \% \textbf{Frequency Discrimination:}
		\STATE \quad Calculate cross-product $ P_{\rm cross}[n]$ via \eqref{eq:cross_prod}
		\STATE \quad Calculate frequency error $\epsilon[n]$ via \eqref{eq:raw_error}
		\STATE \% \textbf{Loop Filtering:}
		\STATE \quad Calculate $\tilde{\epsilon}[n]$ via \eqref{eq:filter_update}
		\STATE \% \textbf{LO Frequency Update:}
		\STATE \quad Calculate cumulative frequency correction: 
		\STATE \quad $\hat{f}_{\text{corr}}[n] = \hat{f}_{\text{corr}}[n-1] + \tilde{\epsilon}[n]$
		\STATE $\quad \text{Update NCO frequency with } \tilde{\epsilon}[n]$
		\STATE $\quad \text{Update physical LO: } f^{\prime}_{\text{LO}}(t) = f_{\text{LO}} + \hat{f}_{\text{corr}}[n]$
		\STATE \quad Update States: $z[n-1], \epsilon[n-1], \tilde{\epsilon}[n-1], \tilde{\epsilon}[n-2]$
		\ENDFOR
	\end{algorithmic}
\end{algorithm}

\section{Simulation Results}
In this section, we evaluate the performance of the proposed adaptive LO tracking Rydberg atomic receiver (RAR) architecture in high-dynamic communication scenarios.

\subsection{Simulation Scenario and Configuration}
We consider the THz high-frequency band communication scenario, where significant Doppler shifts will arise under high mobility. 
To simulate the time-varying Doppler in practical channels, we assume a linear time-dependent Doppler shift $f_d(t)=kt$, where $k$ is the Doppler rate. 
Based on typical setting in non-terrestrial networks \cite{tr38811}, $k$ is set to $816$ kHz/s.

We apply a four-levels ladder-type atomic system with energy levels of $6S_{1/2}, 6P_{3/2}, 20D_{5/2}, 21P_{3/2}$, in which the transition frequency between $20D_{5/2}$ and $21P_{3/2}$ is 309.18GHz, falling within the typical THz band \cite{tajima2016compact}.
Utilizing the ARC library \cite{robertson2021arc}, parameters for atomic receiver, e.g., the dipole moments and decay rates can be directly derived based on the levels.
The detailed parameters are summarized in Table \ref{tab:sim_params}.

\subsection{Simulation Results}

\begin{table}[t]
	\centering
	\caption{Simulation Parameters Settings}
	\label{tab:sim_params}
	\begin{tabular}{lcc}
		\toprule
		\textbf{Parameter} & \textbf{Symbol} & \textbf{Value} \\
		\midrule
		\quad Atom Species & - & $^{85}$Rb \\
		\quad Energy Levels & - & $6S_{1/2}, 6P_{3/2}, 20D_{5/2}, 21P_{3/2}$ \\
		\quad Vapor Cell Length & $L$ & $0.01$ m \\
		\quad Atomic Density & $N_0$ & $4.89 \times 10^{14}$ m$^{-3}$ \\
		\quad Probe Rabi Freq. & $\Omega_p / 2\pi$ & $2.08$ MHz \\
		\quad Coupling Rabi Freq. & $\Omega_c / 2\pi$ & $12.05$ MHz \\
		\midrule
		\quad RF Carrier Frequency & $f_{c}$ & $309.18$ GHz \\
		\quad Intermediate Frequency & $f_{\text{IF}}$ & $1$ MHz \\
		\quad Modulation & - & QPSK \\
		\quad Symbol Rate & $R_s$ & $100$ kBauds \\
		\quad LO Field Amplitude & $A_{\text{LO}}$ & $0.08$ V/m \\
		\midrule
		\quad Natural Frequency & $\omega_n$ & $12500$ rad/s \\
		\quad Damping factor & $\zeta$ & $\sqrt{2}/2$ \\
		\quad Equivalent Gain & $K$ & $1.0$ \\
		\bottomrule
	\end{tabular}
\end{table}

\begin{figure}[t]
	\centering
	\includegraphics[width=0.5\textwidth]{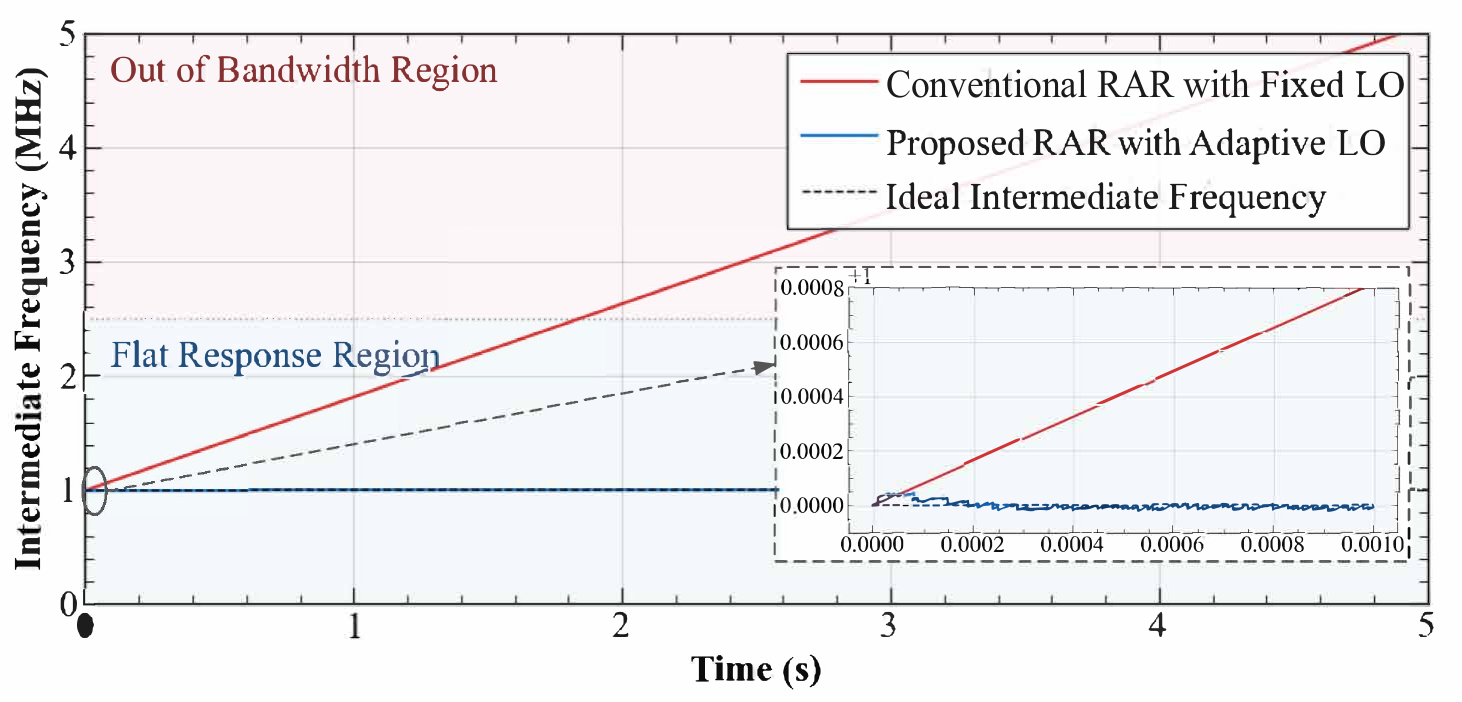}
	\caption{Performance of intermediate frequency tracking.}
	\label{tracking}
\end{figure}

\begin{figure}[t]
	\centering
	\includegraphics[width=0.49\textwidth]{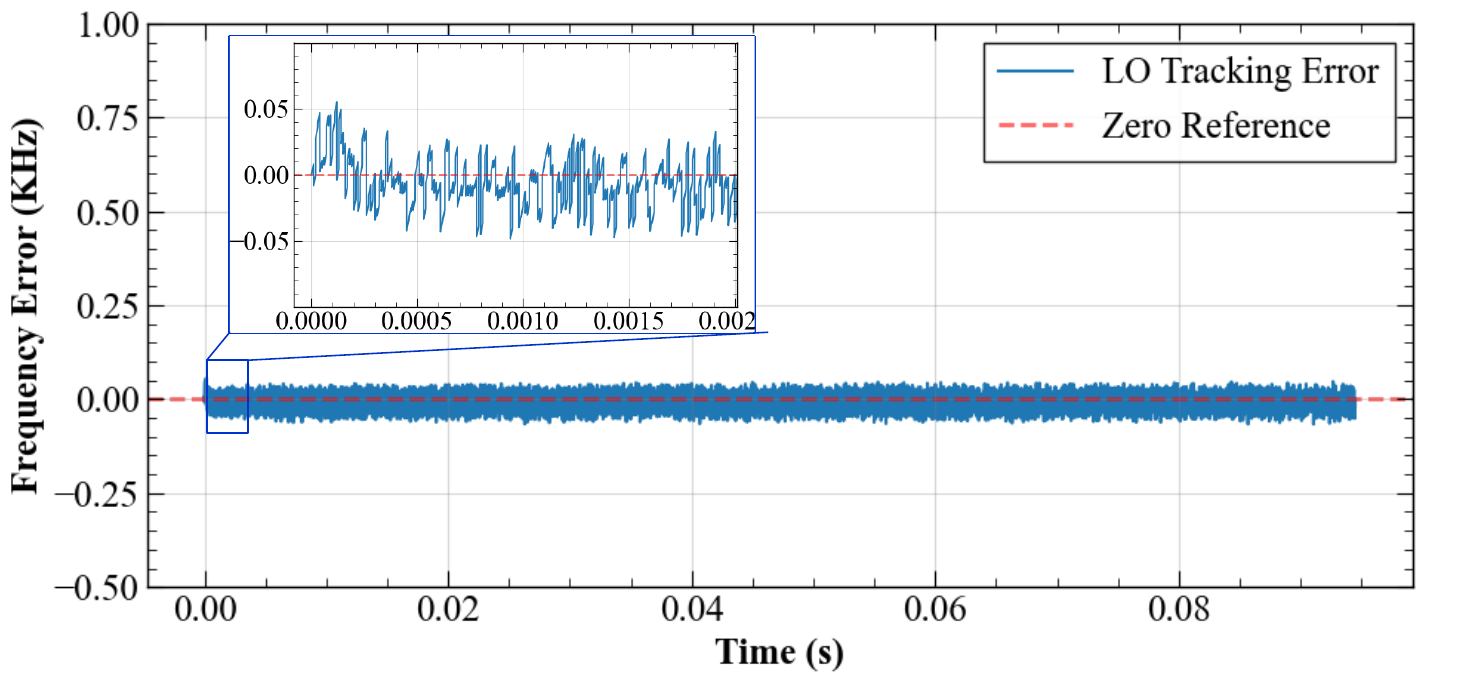}
	\caption{Performance of frequency discrimination error.}
	\label{Error}
\end{figure}

First, we verify the ability to stabilize the effective IF within the atomic flat response region of the proposed architecture. Fig.~\ref{tracking} demonstrates the instantaneous IF over time under Doppler variation. The IF of RAR with fixed LO drifts rapidly over $2.5$MHz and enters the out of bandwidth region in less than $2$ seconds. Compared with the fixed LO, the proposed RAR with adaptive LO tracking locks IF close to $1$MHz, maintaining it within the flat response region all the time.

Fig.~\ref{Error} evaluates the frequency discrimination error of the CPAFC algorithm, to show the tracking performance under the proposed architecture. It can be observed that the frequency discrimination error oscillates closely around zero within $50$Hz over time, revealing that the high precision and stability of the proposed RAR with adaptive LO frequency tracking loop works well in high-dynamic environment.

\begin{figure}[t]
	\centering
	\includegraphics[width=0.5\textwidth]{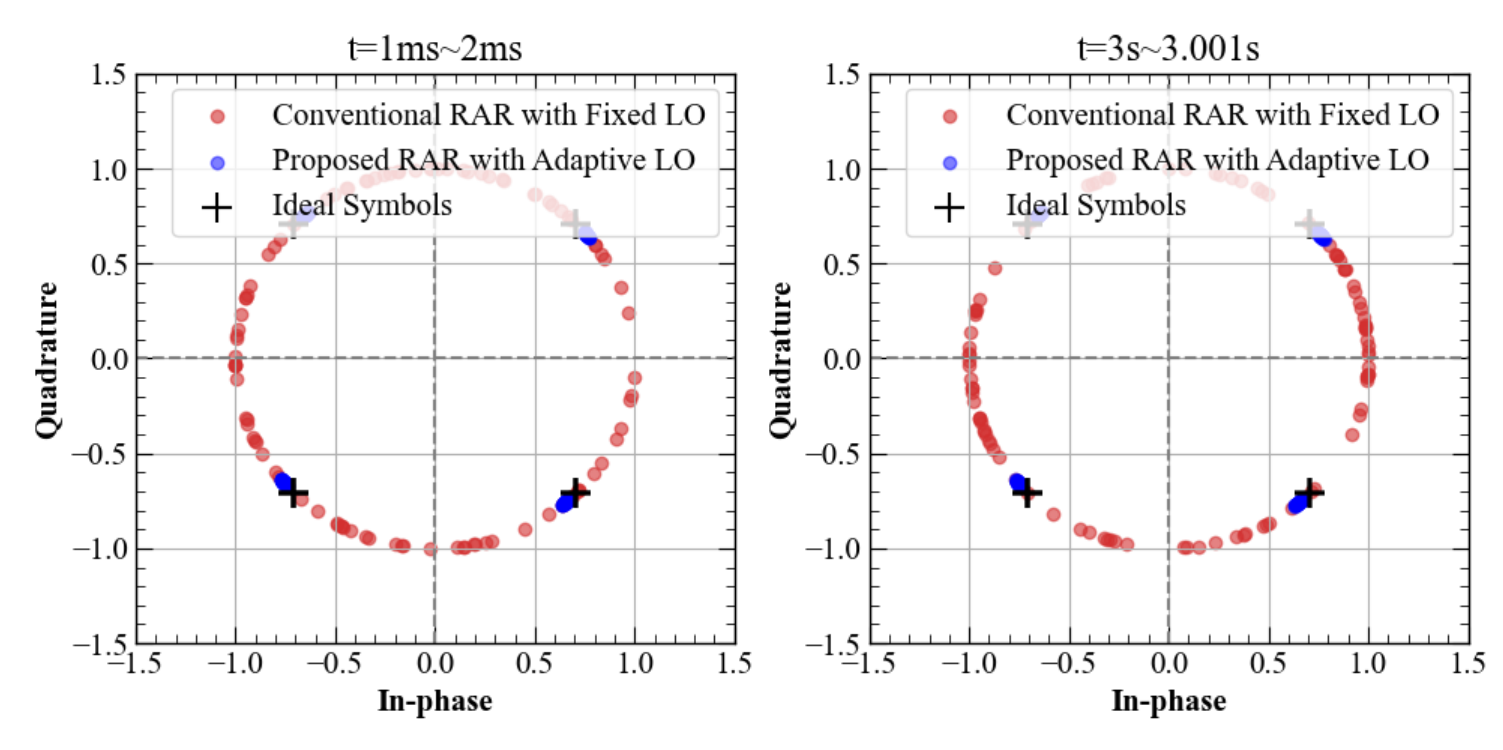}
	\caption{Comparison of constellation @$E_{\rm RF}=0.005V/m$.}
	\label{consellation}
\end{figure}

Fig.~\ref{consellation} compares the received QPSK constellation symbols for the fixed LO scheme and proposed adaptive LO tracking scheme at two distinct time windows, i.e., an early phase of $1\thicksim2$ms and a late phase of $3\thicksim3.001$s. For the fixed LO scheme, severe constellation distortion is observed. The combination of continuous phase rotation and the nonlinear atomic response distortion causes symbols to spread along arcs and eventually form a ring. In contrast, the proposed RAR with LO tracking maintains tightly clustered constellations around all ideal locations at both early and late stages, confirming that the Doppler-induced distortion is effectively mitigated.

\begin{figure}[t]
	\centering
	\includegraphics[width=0.51\textwidth]{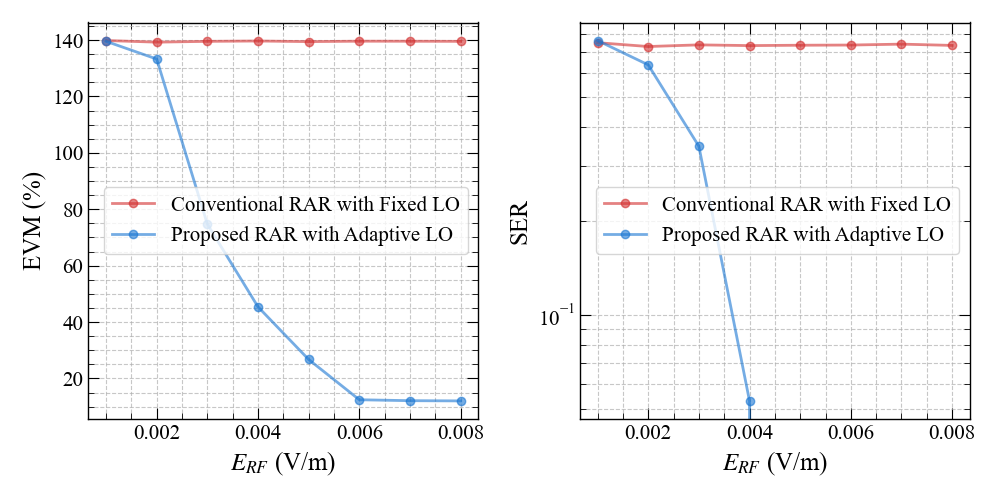}
	\caption{Comparison of EVM and SER @$\sigma^2=1e^{-4}$.}
	\label{EVM&SER}
\end{figure}

Fig.~\ref{EVM&SER} demonstrates the link-level performance across varying RF electric field strengths with fixed moise power $\sigma^2=1e^{-4}$, including error vector magnitude (EVM) and symbol error rate (SER). As shown, the conventional RAR with fixed LO remains at high EVM and SER, while the proposed RAR with adaptive LO tracking decreases significantly with increasing $E_{\rm RF}$, indicating the effectiveness in suppressing signal distortion. and enhancing the overall performance.

%

\section{Conclusion}
This paper proposes an adaptive LO tracking architecture with a frequency locked loop, to addresses the critical challenge of Doppler-induced IF offsets in high-dynamic communications for Rydberg atomic receiver. By employing the CPAFC algorithm, the system estimates instantaneous Doppler shifts and dynamically adjusts the LO frequency, actively locking the effective IF within the atomic response bandwidth to mitigate signal distortion.
Simulation results verify that the proposed receiver alleviates performance degradation under high dynamics. Moreover, the proposed LO tracking architecture is general. Future work may integrate algorithms tailored to specific channel dynamics, signal features or performance requirements, enabling robust Rydberg atomic receiver in practical high-mobility environments.

\bibliography{IEEEabrv,references}  

\end{document}